\documentclass[preprint]{revtex4}
\usepackage{graphicx}

\begin{document}
\begin{center}

{\bf NUCLEAR STRUCTURE \\
AND THE SEARCH FOR COLLECTIVE ENHANCEMENT \\
OF ${\cal P,T}$-VIOLATION}\\

{\bf Naftali Auerbach}$^{1,2}$ and {\bf Vladimir Zelevinsky}$^{2,3}$\\[0.5cm]
{\small $^{1}$School of Physics and Astronomy, Tel Aviv University,
Tel Aviv, 69978, Israel\\
$^{2}$Department of Physics and Astronomy, Michigan State
University, East Lansing, MI 48824-1321, USA\\
$^{3}$National Superconducting Cyclotron Laboratory, Michigan State
University, East Lansing, MI 48824-1321, USA}

\end{center}

\begin{abstract}

The discovery of the atomic electric dipole moment (EDM) would
reveal the simultaneous violation of parity (${\cal P}$) and
time-reversal (${\cal T}$) invariance. The EDM can be induced by
${\cal T}$-odd forces between the nucleus and atomic electrons. As
the nuclear dipole moment is screened according to the Schiff
theorem, the appropriate nuclear operator is the Schiff moment that
may exist in nuclei under ${\cal PT}$-violation. We briefly review
the current experimental situation and discuss more in detail the
ideas concerning possible collective mechanisms for the enhancement
of the nuclear Schiff moment. The most promising directions are
related to the coexistence of octupole and quadrupole collective
modes, either in the form of static deformation or as soft
vibrational excitations. The search for enhancement is important for
widening the pool of nuclei as candidates for the atomic EDM as well
as for development of nuclear many-body theory beyond standard
mean-field and random phase approximations.

\end{abstract}

\section{Introduction}

It became already a common understanding that enables one to claim
that ``the nucleus is a natural laboratory for studying violation of
fundamental symmetries". Indeed, the existence of a broad class of
self-sustaining systems bound by strong interactions, with the
possibility to apply electric and magnetic fields and to observe the
processes governed by weak interactions, allows one to
experimentally select and amplify the phenomena signaling such
violations. The classical example is given by {\sl parity
non-conservation} discovered in the experiment by Wu {\sl et al.}
\cite{wu57} with the use of the polarized $^{60}$Co nucleus. Another
typical feature that characterizes the same case is the
pre-existence of theoretical ideas predicting possible parity
non-conservation \cite{leeyang56}.

Parity non-conservation in nuclear beta-decay is a large effect,
essentially on the maximum possible level (the coefficient of
angular correlation between the beta-electron and nuclear spin is,
as follows from the left-current interaction mechanism,
$\alpha=-v/c$, close to one). Therefore it is currently used as a
tool for extracting important information on aspects of nuclear
structure unrelated to weak interactions. As an example, we can
mention the developing idea \cite{horowitz01} of utilizing the
interference of electromagnetic and weak interactions for a
measurement of the neutron skin in heavy nuclei (the ``weak charge"
of the proton in the standard model is only 0.08 of the neutron weak
charge). Such data, still unavailable, would be instrumental for
information on nuclear symmetry energy \cite{brown00} and the
equation of state of neutron matter \cite{piekarewicz08}.

In general, the main interest to violation of fundamental symmetries
can be explained by the search for the effects beyond the
predictions of the standard model \cite{erler05}. Among other
problems, such violations can provide a key to the understanding of
one of the critical scientific puzzles, the baryon asymmetry of the
universe. One of the famous Sakharov conditions for baryogenesis of
asymmetry \cite{sakharov67} is the ${\cal CP}$-violation, which is
equivalent, according to the ${\cal CPT}$-theorem, to the violation
of the time-reversal (${\cal T}$) invariance. The small effects of
${\cal CP}$- violation known from experiments with neutral $K$- and
$B$-mesons \cite{fleischer02} set limits on the deviations from the
standard model. We still do not have numbers which would show the
strength of ${\cal CP}$-violation in nucleon-nucleon and/or
quark-quark forces. Our best hopes in this direction are connected
with atomic and nuclear experiments \cite{ginges04}.

In contrast to the beta decay, here we are interested in the
specific properties of minor components of {\sl stationary} atomic
and nuclear states. The non-orthodox features can be better seen in
phenomena fully forbidden by standard physics. Since anyway these
effects are numerically small, the main idea is to use the atomic
or/and nuclear environment as a possible {\sl amplifier} of
non-standard properties. The experience of last decades confirms
that this idea may work.

As a striking example we can remind the reader the strong enhancement of the
parity violation in scattering of slow longitudinally polarized neutrons off
heavy targets first measured in Dubna \cite{alfimenkov83} and then studied in
detail at Los Alamos, see review articles
\cite{bowman93,frankle93,flamgrib95,mitchell99} and references therein. The
relative difference of cross sections for neutrons of right and left
longitudinal polarization can reach 10\%, whereas simple estimates based on a
typical strength of weak interactions would give $10^{-(7-8)}$. The enhancement
by six orders of magnitude comes from the combination of high density of
$p$-wave and $s$-wave resonances, uniformly chaotic wave functions of compound
states, and the large kinematic ratio of neutron widths $\Gamma_{s}/\Gamma_{p}$
for the resonances mixed by ${\cal P}$-violating forces. This enhancement was
anticipated long ago \cite{blin60}, theoretically predicted before the
measurements by Sushkov and Flambaum \cite{SF80,SF82} and discussed in detail
in \cite{ABowman92,ABAB94}.

A related, and probably even less expected, enhancement by 3-4
orders of magnitude was observed in the asymmetry of the fragments
from nuclear fission by slow polarized neutrons with respect to the
neutron polarization, see the review of the first experiments in
\cite{danilyan80}. Here the explanation \cite{SF82} again is based
on a chaotic character of intrinsic nuclear dynamics \cite{big,ann}.
The kinematic enhancement factor is absent but the new element is
that, after the weak interaction has mixed compound levels of
opposite parity, the pear-shaped nucleus with parity doublets (an
analog of $\Lambda$-doubling in molecules \cite{LLQM}) proceeds
through few specific fission channels which preserve memory of
parity non-conservation. The later Grenoble experiments
\cite{kotzle00} confirmed this scenario showing that the mixing
occurred in a ``hot" nucleus between chaotic wave functions, and
therefore the resulting asymmetry is practically independent on the
mass and kinetic energy distributions determined at the next stages
of the process.

We mentioned above the impressive nuclear phenomena related to
parity non-conservation. The ${\cal T}$-violation was searched in
various nuclear reactions \cite{blinbook} but also in nuclear
structure, namely in chaotic regions of complex spectra of excited
nuclear states \cite{french88}. The character of level repulsion at
small spacings critically depends on the presence or absence of
time-reversal invariance of the Hamiltonian: the probability of
having two levels of the same symmetry class at a small spacing $s$
behaves as $s$ and $s^{2}$ with and without ${\cal T}$-invariance,
respectively. Because of poor statistics for closely spaced avoided
level crossings, in practice such analysis produces only a rough
upper boundary. The most interesting and promising direction is
related to the experimental search for electromagnetic multipoles
forbidden by discrete symmetries. As explained in the next section,
the measurement of a non-vanishing expectation value $\langle {\bf
d}\rangle$ of the electric dipole moment (EDM) ${\bf d}$ of the atom
would be a direct proof of the existence in atoms and nuclei of the
interactions violating {\sl simultaneously} ${\cal P}$- and ${\cal
T}$-symmetry \cite{purcell50,landau57}. The search for the atomic
EDM is going on in few experimental groups; right now the best
boundaries are established for \cite{jacobs95,romalis01} for
$^{129}$Xe and $^{199}$Hg. Our short review is devoted to the
discussion of nuclear dynamics which could amplify the EDM and
therefore facilitate those hard and time-consuming experiments.

The ideas of possible nuclear enhancement of the atomic effects are
important for choosing the right nuclear isotope with largest odds
of the successful measurement. In contrast to the statistical
enhancement of parity non-conservation in neutron scattering, the
EDM is to be found in the ground state of the atom and the nucleus.
Here one cannot hope to profit from the uniformly chaotic states at
high level density. The chances to get significant enhancement are
related either to the more or less accidental close proximity of
levels capable of being mixed by the weak interactions
\cite{haxton83}, or with possible collective effects which could
{\sl coherently} enhance the atomic EDM. In the EDM problem, the
desired effects might be associated with the combination of the
quadrupole and octupole collectivity, either in the form of static
deformation, or in the interplay of corresponding soft vibrational
modes. These ideas will be discussed in detail below. Apart from the
direct goal to find the source for the enhancement of weak
interactions, such studies are interesting by itself, as a many-body
problem going beyond standard mean-field and random phase
approximations.

We start with the explanation,  Sec. 2, of the role of the nuclear
Schiff moment as a mediator of ${\cal P,T}$-odd forces between the
nucleus and atomic electrons. We also briefly go through the list of
current experimental approaches to search for the EDM. A simple
single-particle contribution to the Schiff moment is discussed in
Sec. 3. Sec. 4 describes the enhancement that is possible in heavy
nuclei under simultaneous presence of static quadrupole and octupole
deformation. In Sec. 5 we proceed with the further ideas
substituting the permanent deformation with the large-amplitude
vibrations of corresponding multipolarities. The appropriate
conditions may exist in radioactive isotopes of radium, radon and
thorium. These ideas are more speculative since the fully realistic
calculations were not performed yet. Sec. 6 contains a short
summary.

\section{EDM and Schiff moment}

\subsection{EDM and fundamental symmetries}

The electric dipole moment of the static charge distribution in the
atom is given by the expectation value of the operator $\hat{{\bf
d}}=\sum_{a}e_{a}{\bf r}_{a}$ in the ground state of the ato, where
the subscript $a=1,...,Z$ enumerates atomic electrons. The dipole
operator is a polar, i.e. ${\cal P}$-odd, ${\cal T}$-even, vector.
In order to have a non-zero expectation value of any vector, the
stationary state must have non-zero angular momentum. We will denote
the total atomic spin ${\bf J}$ and the nuclear spin ${\bf I}$. This
excludes closed shell atoms and nuclei. Nuclear spin $I\neq 0$
requires an odd mass number. In fact, the nuclei of current
experimental interest, such as $^{129}$Xe, $^{199}$Hg, and
$^{225}$Ra, have $I=1/2$ which might be advantageous since the
time-reversed magnetic substates $M_{I}=\pm 1/2$ are not split by
external electric fields.

The rotational invariance supposedly remains exact even in the
presence of weak interactions. This implies the well known relations
between the matrix elements of any vector and those of angular
momentum. Inside the rotational multiplet of states $|JM\rangle$,
the EDM acts as an effective operator
\begin{equation}
\hat{{\bf d}}=\,\frac{\langle {\bf d}\cdot{\bf J}\rangle}
{J(J+1)}\,\hat{{\bf J}};                            \label{1}
\end{equation}
here and below we mark by a hat the quantities where it is important
to stress the operator nature. If the stationary state under study
has definite parity, the expectation value $\langle {\bf d}\cdot{\bf
J}\rangle$ vanishes being the pseudoscalar product of a polar and an
axial vector. Obviously, the existence of the EDM necessarily
requires ${\cal P}$-violation.

However, this is not sufficient. The rotational scalar $\langle {\bf
d}\cdot{\bf J}\rangle$ cannot depend on the projection $J_{z}$ of
the atomic state; therefore it should have the same value for
$J_{z}=M$ and $J_{z}=-M$. On the other hand, the transformation from
$M$ to $-M$ is equivalent to time reversal, when ${\bf J}$ changes
sign while ${\bf d}$ does not. Therefore the non-zero expectation
value of any polar ${\cal T}$-even vector requires, in addition, a
violation of ${\cal T}$-invariance (the classical {\sl
Purcell-Ramsey theorem} \cite{purcell50}). In contrast to that, the
non-zero expectation values of ${\cal P}$-even ${\cal T}$-odd vector
operators, such as the magnetic moment, do not require any violation
of fundamental symmetries. The arguments do not change in the
presence of the hyperfine structure, one just needs to substitute
the total angular momentum ${\bf F}={\bf J}+{\bf I}$ instead of the
atomic spin ${\bf J}$.

\subsection{Schiff theorem}

In order to see how the ${\cal P,T}$-violating hadronic forces
induce the atomic EDM we consider (for simplicity, in the
non-relativistic approximation) the Hamiltonian of the atomic
system,
\begin{equation}
H_{{\rm atom}}=H_{{\rm el}}+ H_{{\rm nucl}}+\sum_{a}e\phi({\bf
r}_{a}),                                           \label{2}
\end{equation}
where $H_{{\rm el}}$ and $H_{{\rm nucl}}$ are total Hamiltonians of
interacting electrons and nucleons, respectively, while $\phi({\bf
r})$ is the electrostatic potential generated by the nuclear ground
state charge density $\rho({\bf x})$,
\begin{equation}
\phi({\bf r})=\,\int d^{3}x\,\frac{\rho({\bf x})}{|{\bf r}-{\bf
x}|}.                                          \label{3}
\end{equation}
The EDM is probed by the external electric field ${\bf E}$ that
interacts with electrons and protons,
\begin{equation}
H_{{\rm ext}}=-\Bigl(e\sum_{a}{\bf r}_{a}+\hat{{\bf D}}\Bigr)
\cdot{\bf E},                                   \label{4}
\end{equation}
where $\hat{{\bf D}}$ is the operator of the nuclear electric dipole
moment. Due to ${\cal PT}$-violation, this operator can have a
non-zero expectation value $\langle{\bf D}\rangle$ in the exact
nuclear ground state.

In order to proceed with the calculation of the response of the
system to the electric field, we follow the method of Refs.
\cite{SAF97} (see the Appendix) and \cite{reexamschiff}. It is
convenient to make a canonical transformation of the full
Hamiltonian ${\cal H}=H_{{\rm atom}}+H_{{\rm ext}}$ using a unitary
operator $(\hbar=1)$
\begin{equation}
\hat{U}=\,\frac{\langle{\bf D}\rangle}{Z|e|}\cdot\sum_{a}\hat{{\bf
p}}_{a}.                                        \label{5}
\end{equation}
We can limit ourselves to the first order with respect to the small
quantities like $\langle{\bf D}\rangle$, so that
\begin{equation}
{\cal H}'=e^{i\hat{U}}{\cal H}e^{-i\hat{U}}\approx {\cal
H}+i[\hat{U},{\cal H}].                        \label{6}
\end{equation}
The commutator in eq. (\ref{6}) is given by
\begin{equation}
i[\hat{U},{\cal H}]=\frac{1}{Z}\langle{\bf D}\rangle\cdot(Z{\bf E}
+{\bf E}_{e}),                                        \label{7}
\end{equation}
where
\begin{equation}
{\bf E}_{e}=-\sum_{a}\nabla_{a}\phi({\bf r}_{a})     \label{8}
\end{equation}
can be interpreted as the electric field on the nucleus produced by
atomic electrons. In a stationary state $|\Psi\rangle$ of the full
Hamiltonian ${\cal H}$,
\begin{equation}
\langle \Psi|[\hat{U},{\cal H}]|\Psi\rangle=0.        \label{9}
\end{equation}
This means that the stationary state in the external field polarizes
the electron configuration in such a way that the total field acting
on the nucleus vanish,
\begin{equation}
Z{\bf E}+\langle {\bf E}_{e}\rangle=0.              \label{10}
\end{equation}
This screening is the content of the Schiff theorem
\cite{schiff63,sandars67}. In simple terms, in a stationary state,
forces on the nuclear dipole are to be compensated. Now we need to
consider what is left in the Hamiltonian after the canonical
transformation (\ref{6}).

The interaction of the external field with the nuclear dipole is
mostly taken in account, except for the fluctuational term $-({\bf
D}-\langle{\bf D}\rangle)\cdot{\bf E}$. Therefore in the first order
the external field does not lead to the energy shift (in higher
orders it still influences the nuclear polarizability). The external
field however renormalizes the electron-nucleus interaction: instead
of the usual electrostatic potential $\phi({\bf r})$ we have now
\begin{equation}
\phi'({\bf r})=\phi({\bf r})-\frac{1}{Ze}\,\langle{\bf D}\rangle
\cdot\nabla\phi({\bf r}).                        \label{11}
\end{equation}
This is the starting point of the path that leads to the Schiff
moment.

We have to mention that a different form of the canonical
transformation,
\begin{equation}
\tilde{U}=\,\frac{{\bf D}}{Z|e|}\cdot\sum_{a}\hat{{\bf p}}_{a},
                                                   \label{12}
\end{equation}
was considered recently \cite{liu07}. Here the operator $\hat{{\bf
D}}$ is used, in contrast to its expectation value in our version
(\ref{5}). The transformation (\ref{12}), leading to the exact
cancelation instead of our fluctuation term, brings instead the
commutators of the operator $\hat{{\bf D}}$ with the full nuclear
Hamiltonian and therefore introduces complicated nuclear
correlations. The derivation above is in agreement with previous
results for the Schiff moment.

As indicated in the pioneering work by Schiff \cite{schiff63}, the
screening theorem is violated by the hyperfine interactions.
Actually, this effect turns out to dominate in hydrogen and helium
atoms \cite{dzuba07}. However, in heavy atoms, which are subject of
our main interest, the contribution of the Schiff moment (see the
next subsection), together with the relativistic enhancement of the
electronic wave functions in the vicinity of the nucleus, makes the
situation much more promising.

\subsection{Schiff moment}

The standard expression for the Schiff moment \cite{FKS84} can be
derived in many ways. If we neglect the relativistic corrections of
the order $(Z\alpha)^{2}$ we can use a simple expansion of the
nuclear charge density $\rho({\bf r})$ in the series over gradients
of the delta-function,
\begin{equation}
\rho({\bf x})=\left\{A+({\bf B}\cdot\nabla)+
\frac{1}{2}C_{ik}\nabla_{i}\nabla_{k}+\dots\right\}\delta({\bf x}).
                                               \label{13}
\end{equation}
A more precise derivation accounting for relativistic corrections
can be found in \cite{flamginges02,dmitriev05}. The coefficients of
the expansion (\ref{13}) are related to the nuclear multipole
moments,
\begin{equation}
\int d^{3}x\,\rho({\bf x})=Z|e|,                 \label{14}
\end{equation}
\begin{equation}
\int d^{3}x\,\rho({\bf x}){\bf x}=\langle {\bf D}\rangle, \label{15}
\end{equation}
\begin{equation}
\int d^{3}x\,\rho({\bf x})(3x_{i}x_{k}-\delta_{ik}{\bf
x}^{2})=Z|e|\langle Q_{ik}\rangle,
\label{16}
\end{equation}
\begin{equation}
\int d^{3}x\,\rho({\bf x})x^{2}=Z|e|\langle x^{2}\rangle_{{\rm ch}},
                                                      \label{17}
\end{equation}
\begin{equation}
\int d^{3}x\,\rho({\bf x})x^{2}{\bf x}=\langle {\bf D}^{(2)}\rangle,
                                                     \label{18}
\end{equation}
etc. The operator ${\bf D}^{(2)}$ is associated \cite{harakeh81}
with the isoscalar dipole giant resonance since the usual isoscalar
dipole moment reduces to the center-of-mass excitation.

Expressing the charge density (\ref{13}) in terms of physical
mutlipoles, we obtain
\begin{equation}
\rho({\bf x})=\sum_{l}\rho^{(l)}({\bf x})\delta({\bf x}), \label{19}
\end{equation}
where the multipole operators $\rho^{(l)}$ acting on the
delta-function are given by
\begin{equation}
\rho^{(0)}=Z|e|\left\{1+\frac{1}{6}\,\langle x^{2}\rangle_{{\rm ch}}
\nabla^{2}+\,\dots \right\},
\label{20}
\end{equation}
\begin{equation}
\rho^{(1)}=-\left(\langle {\bf D}\rangle+\frac{1}{10}\langle{\bf
D}^{(2)}\rangle \nabla^{2}\right)\cdot\nabla+\,\dots, \label{21}
\end{equation}
\begin{equation}
\rho^{(2)}=\frac{Z|e|}{6}\,\langle
Q_{ik}\rangle\nabla_{i}\nabla_{k}+\,\dots
                                                     \label{22}
\end{equation}

This expansion enters the effective potential $\phi'({\bf r})$ of
eq. (\ref{11}) that, after taking the expectation value in the
nuclear ground state $|0\rangle$, becomes
\begin{equation}
\langle 0|e\phi'({\bf r})|0\rangle=-\frac{Ze^{2}}{r}+4\pi e({\bf
S}\cdot \nabla)\,\delta({\bf r})+\,\dots, \label{23}
\end{equation}
where ${\bf S}$ is the expectation value of the Schiff moment vector operator,
\begin{equation}
\hat{S}_{i}=\frac{1}{10}\int d^{3}x\,\rho({\bf x})\left\{x^{2}x_{i}
-\frac{5}{3} \langle x^{2}\rangle_{{\rm ch}}x_{i}-\frac{2}{3}\langle
Q_{ik}\rangle x_{k}\right\}.                        \label{24}
\end{equation}
In addition, this operator can have contributions from the possible
internal dipole moments of the nucleons \cite{ginges04}. These
contributions cannot be experimentally distinguished from those
determined by the nuclear multipoles in eq. (\ref{24}).

\subsection{From nuclear Schiff moment to atomic EDM}

The expectation value of the nuclear Schiff moment (\ref{24}),
similar to the atomic EDM (\ref{1}), is given by the effective
operator of the vector model for the nuclear ground state with spin
${\bf I}>0$,
\begin{equation}
\hat{\bf S}=\frac{\langle{\bf S}\cdot{\bf I}\rangle}{I(I+1)}\,
\hat{{\bf I}}.                                        \label{25}
\end{equation}
The exact nuclear state $|I\rangle$ should be a superposition of the
ground state $|I;0\rangle$ (found in the absence of weak
interactions) and admixtures of the opposite parity states
$|I;k\rangle$ induced by the ${\cal PT}$-violating weak interaction
$W$,
\begin{equation}
|I\rangle=|I;0\rangle+\sum_{k\neq 0}\frac{\langle I;k|W|I;0\rangle}
{E_{0}-E_{k}}.                                 \label{26}
\end{equation}
The weak perturbation (\ref{26}) creates the non-zero expectation
value of the Schiff moment,
\begin{equation}
\langle {\bf S}\rangle=2\,{\rm Re}\,\sum_{k\neq 0}\frac{\langle I;0|{\bf
S}|I;k\rangle\,\langle I;k|W|I;0\rangle}{E_{0}-E_{k}}. \label{27}
\end{equation}

The unknown ${\cal PT}$-violating weak interaction $W$, under the
assumption of two-body nucleon-nucleon forces and to the first order
in the nucleon velocities, can be parameterized as \cite{FKS84}
\[W_{ab}=\frac{G}{\sqrt{2}}\,\frac{1}{2m}\,\{(\eta_{ab}
\vec{\sigma}_{a}-\eta_{ba}\vec{\sigma}_{b})\cdot\nabla_{a}\delta({\bf
x}_{a}-{\bf x}_{b})\]
\begin{equation}
+\eta'_{ab}[\vec{\sigma}_{a}\times\vec{\sigma}_{b}][({\bf p}_{a}-
{\bf p}_{b}),\delta({\bf x}_{a}-{\bf x}_{b})]_{+}\}.    \label{28}
\end{equation}
Here $G$ is the Fermi constant of the weak interaction, $m$ is the
nucleon mass, while $\vec{\sigma}_{a,b},\; {\bf x}_{a,b},$ and ${\bf
p}_{a,b}$ are the spins, positions and momenta, respectively, of the
interacting nucleons $a$ and $b$; $[,]_{+}$ means an anticommutator.
The dimensionless coupling constants, $\eta_{ab}$ and $\eta'_{ab}$,
of the ${\cal P,T}$-violating forces are to be extracted from the
values of the atomic EDM. Theoretical arguments
\cite{khatsymovsky88,falk99} show that such an interaction is
determined mainly by the exchange of neutral pions.

The Schiff moment influences the atomic electrons through the second
term of the potential (\ref{23}); a more accurate account of the
finite size of the nucleus can be found in Ref. \cite{flamginges02}.
This mixes electron orbitals of opposite parity and generates the
EDM of the atom. The result of detailed atomic calculations
\cite{dzuba02} can be presented as the proportionality between the
atomic EDM $d$ and the Schiff moment $S$, eq. (\ref{27}),
\begin{equation}
d=\xi\,\left(\frac{S}{e\cdot{\rm fm}^{3}}\right)10^{-17}e\cdot{\rm cm}.
                                                \label{29}
\end{equation}
The  numerical factor $\xi$ is the main result of complicated calculations,
which can be a topic of a special review article. This factor grows with the
nuclear charge reaching the values 3.3 for radon and -8.5 for radium.

\subsection{Briefly on the experimental situation}

As stated in the first line of Ref. \cite{jacobs95}, ``No permanent
electric-dipole moment (EDM) of an elementary particle, atom, or
molecule has yet been detected after several decades of
experimentation". We are interested here only in the experiments
which are sensitive to the nuclear spin and Schiff moment. Presently
the measurement of the atomic EDM is pursued by several experimental
groups, and we mention only few recent results based on various
cutting-edge applications of methods of quantum optics. There exist
also interesting suggestions for the EDM measurement in storage
rings \cite{khriplovich98,farley04,orlov06}.

The data obtained for $^{129}$Xe and $^{199}$Hg provide the best
available limits. The measurement \cite{rosenberry01} of the Zeeman
splitting in parallel or antiparallel electric and magnetic fields
with simultaneous presence of laser polarized $^{129}$Xe and
$^{3}$He (the latter served as a ``comagnetometer") gave
$d(^{129}{\rm Xe}=(0.7\pm 2.8)\times 10^{-27}\,e$ cm. The upper
boundary value for the EDM of $^{199}$Hg was given in Ref.
\cite{jacobs95} as $|d(^{199}{\rm Hg})|<8.7\times 10^{-28}e\cdot$cm
which was better by a factor of 25 than the previous mercury
measurement \cite{lamoreaux87}. The method of \cite{jacobs95}
employed the measurement of the spin precession in the electric
field and had many technical improvements compared to
\cite{lamoreaux87}. The next step in the same direction was made by
the same group at the University of Washington in Ref.
\cite{romalis01}. By using a different technique of measuring the
Zeeman precession of  nuclear spins in parallel electric and
magnetic fields, the limit for the EDM of $^{199}$Hg was lowered by
the factor of 4 up to $|d(^{199}{\rm Hg})|<2.1\times
10^{-28}e\cdot$cm. Already this limit puts stringent constraints on
${\cal CP}$-violating effects beyond the standard model
\cite{romalis01}.

The results of the Berkeley experiment \cite{regan02} performed on
atomic $^{205}$Tl in the ground state with the use of the
atomic-beam magnetic resonance and laser optical pumping improved
the results of the previous attempts \cite{commins94} and were
interpreted as a limit for the electric dipole of the electron,
$|d_{e}|\leq 1.6\times 10^{-27}e\,$cm. Essentially, here the signal
of ${\cal P,T}$-violation is given by the dependence on the ${\cal
P,T}$-odd relativistic invariant $({\bf E}\cdot{\bf B})$ of
electromagnetic fields.

Isotopes of radium and radon seem to be the appropriate candidates
for the combination of nuclear and atomic enhancement factors; we
will discuss later more in detail the advantages of heavy nuclei
with a combination of quadrupole and octupole deformation. Among
these isotopes $^{225}$Ra is one of the most attractive due to its
nuclear spin 1/2 and reasonably long lifetime $t_{1/2}=15$ days.
Specific near-degeneracies in the atomic spectrum can lead to
further enhancements \cite{dzuba00,dzuba02}. Because of this
importance, the success of the Argonne group \cite{guest07} can be
estimated as a sign of the serious progress in this direction. It
turned out to be possible to perform laser trapping of $^{225}$Ra
(as well as $^{226}$Ra)  and to measure the isotope shift, hyperfine
splittings and lifetimes of certain levels in cooled atoms. The
thermal black-body radiation at room temperature served as an
instrument in the redistribution of level populations.

Lighter radium and radon isotopes, which also could, as we discuss
below, compete for favorable Schiff moment conditions, were not
sufficiently studied until now. They mostly have much shorter
alpha-decay lifetime (only $^{223}$Ra has $t_{1/2}=11$ days but
nuclear spin $I=3/2$ and the non-zero quadrupole moment that may
lead to systematic errors in atomic experiments). The successful
techniques of nuclear orientation of radon isotopes $^{209}$Rn and
$^{223}$Rn by spin-exchange optical pumping was developed long ago
\cite{kitano88}. An experiment to measure the atomic EDM of
$^{223}$Ra is planned at TRIUMF (E-929 collaboration) with results
expected in a number of years (the production rate of $^{223}$Rn at
the ISAC facility is $10^{7}$/s). Also the KVI group is planning to
study the EDM of laser-trapped radium isotopes, in particular of
$^{225}$Ra. A more detailed information on the last two works, as
well as about the experimental situation in general, can be found on
line in the presentations by T. Chupp and K. Jungman at the workshop
at INT, Seattle, 2007 \cite{INT07}. In the light of ideas of
possible soft mode enhancement the pool of nuclear candidates can
become wider. There are also ideas in the literature of using
molecular and condensed matter systems with strong local electric
fields \cite{hudson02,lamoreaux02,mukhamedjanov05} where experiments
are under preparation, see for example, presentation by D. Budker
\cite{INT07}.

\section{Single-particle Schiff moment}

Now we concentrate on the magnitude of the nuclear Schiff moment
that results from complicated many-body dynamics in heavy nuclei. In
the simplest meaningful approximation, the ground state of an
odd-$A$ nucleus has one unpaired particle. In this picture the
many-body dynamics is reduced to the mean field that defines the
symmetry of single-particle motion. This is what is usually called
{\sl nuclear shape} \cite{BM}. Another important ingredient is the
pairing interaction that introduces the {\sl seniority} quantum
number $v$ which can be identified with the number of unpaired
particles. In this approximation $v=0$ for the ground state of an
even-even nucleus, and $v=1$ in the odd-$A$ case.

\begin{figure}
\begin{center}
\includegraphics[width=12cm]{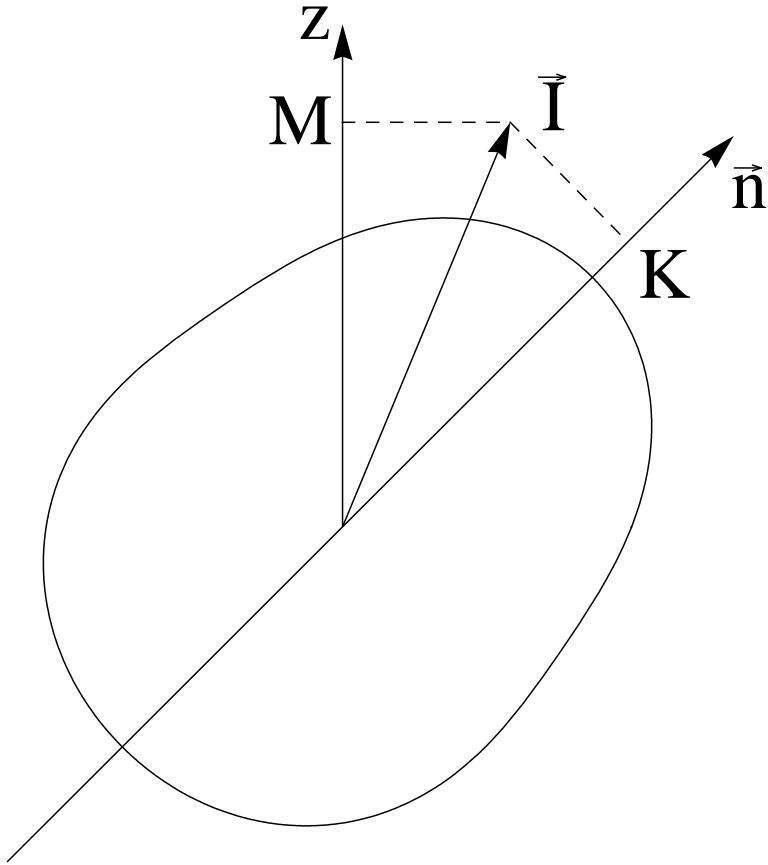}
\caption{The intrinsic shape corresponding to the axially symmetric
combination of quadrupole and octupole deformation; parameters are
$\beta_{2}=0.3$ and $\beta_{3}=0.1$; the angular momentum vector
${\bf I}$ has the body-fixed projection $K$ and the space-fixed
projection $M$. \label{Puc1}}
\end{center}
\end{figure}

The shape is defined with respect to the intrinsic ({\sl
body-fixed}) reference frame, see the conventional Fig. 1. This does
not matter in the {\sl spherical} case, when the mean field and
spin-orbit coupling generate single-particle orbitals with definite
angular momentum ${\bf j}={\bf l}+{\bf s}$ and degeneracy with
respect to the arbitrarily chosen projection $j_{z}=m$. However, in
the {\sl deformed} case, the reference frame is rotating together
with the body, and this rotational motion defines the orientational
wave function $D^{I}_{MK}$, where $I$ and $I_{z}=M$ are total
nuclear spin and its projection onto the laboratory quantization
axis. Here and below we assume that the shape has {\sl axial
symmetry} characterized by the unit vector ${\bf n}$ and one can
define the conserved component $({\bf I}\cdot{\bf n})=K$ of the
total spin along the symmetry axis. The single-particle states in
this case are Nilsson orbitals with a certain angular momentum
projection $j_{z}=\kappa$. Due to time-reversal invariance of strong
forces, the orbitals $\pm\kappa$ are degenerate ({\sl Kramers
theorem}). The presence of deformation of various multipolarities is
an important resource in search for the enhancement of violation of
fundamental symmetries.

In the body-fixed frame there are no angular momentum restrictions,
and any multipole operator, including the Schiff moment, can have a
non-zero intrinsic value. The observable value of ${\bf S}$ in the
space-fixed frame appears only due to the mixing induced by the weak
interaction. It is believed that the main contribution comes from
the coherent mean-field part of the interaction (\ref{28}) that can
be written as a one-body operator,
\begin{equation}
\overline{W}({\bf x})=\frac{G}{\sqrt{2}}\,\frac{1}{2m}\,\frac{\eta}{4\pi}\,
(\vec{\sigma}\cdot\nabla)\rho({\bf x}),                   \label{30}
\end{equation}
where $\rho({\bf x})$ is the nuclear density. In the single-particle
approximation, the weak interaction mixes orbitals of opposite parity; in the
spherical case they have $j'=j$ but different orbital momenta, $l'=l\pm 1$.

Rare accidental proximity of the orbitals of opposite parity can
lead \cite{haxton83} to the small energy denominators in (\ref{27}).
However, for single-particle mixing, this can hardly enhance the
outcome since the matrix elements of $\overline{W}$, eq. (\ref{30}),
are roughly proportional to those of the single-particle momentum
operator, and therefore to the energy differences of the mixed
orbitals \cite{FKS84}. This cancels the small energy denominators.

The next step beyond the single-particle model should account for
the residual interaction (pairing is in a sense a part of the
mean-field picture). Realistic calculations were performed in
various versions of configuration mixing
\cite{FKS86,FV94,DS03,DSA05,jesus05}. The main effect appearing here
is the core polarization by the unpaired particle. The
quasiparticles are dressed by this polarization and all observables
are renormalized. In agreement with the previous experience, the
renormalization can lead to qualitatively new results if there exist
low-lying {\sl collective modes} of considerable strength. But we do
not know about significant collective dipole strength in the
vicinity of the ground state. Because of this absence of coherence,
the results of such calculations for the Schiff moment differ from
pure single-particle estimates not more than by the factor of about
2.

\section{Static deformation}

We have to search for specific structural features which can bring
closely such levels of opposite parity that can have a large
probability of being mixed by the interaction $W$. These features
are related to the possible {\sl coherence} of mixing that involves
collective contributions of many particles. The main attempts in
this direction use {\sl deformed} nuclei as the appropriate arena
for the combined action of intrinsic symmetry and weak interactions.

\subsection{Combination of quadrupole and octupole deformation}

Let us consider an axially symmetric deformed odd-$A$ nucleus. The deformed
mean field spontaneously breaks rotational symmetry. In the usual
approximation, the nuclear rotation (which plays the role of the Goldstone mode
that restores the proper quantum numbers of angular momentum) is adiabatic with
respect to intrinsic excitations. The full wave function can be presented as
the product of the rotational Wigner function $D^{I}_{MK}$ depending on the
orientational angles and the intrinsic function $\chi_{K}$, where $K$ is an
intrinsic pseudoscalar. In the body-fixed frame, any polar vector, such as the
Schiff moment ${\bf S}$, can have a non-zero expectation value ${\bf S}_{{\rm
intr}}$ without any ${\cal P}$- or ${\cal T}$-violation. The symmetry dictates
the direction of this vector along the symmetry axis,
\begin{equation}
{\bf S}_{{\rm intr}}=S_{{\rm intr}}{\bf n}.           \label{31}
\end{equation}
However, this intrinsic vector is averaged out by rotation because
the only possible combination in the space-fixed frame is again
similar to the one we have seen in eqs. (\ref{1}) and (\ref{25}),
namely proportional to the pseudoscalar product $\langle({\bf
n}\cdot {\bf I})\rangle$ that violates ${\cal P}$- and ${\cal
T}$-invariance. If the ${\cal P,T}$-violating forces create an
admixture $\alpha$ of states of the same spin and opposite parity,
the average orientation of the nuclear axis arises. In the linear
approximation with respect to $\alpha$,
\begin{equation}
\langle ({\bf n}\cdot{\bf I})\rangle=2\alpha K,       \label{32}
\end{equation}
and, therefore, we acquire the space-fixed Schiff moment ({\ref 25}) along the
laboratory quantization axis,
\begin{equation}
\langle IM|\hat{{\bf S}}|IM\rangle=S_{{\rm intr}}\,\frac{2\alpha KM} {I(I+1)}.
                                                         \label{33}
\end{equation}
Now the idea is to obtain a large intrinsic Schiff moment and not to lose much
in translating the result to the space-fixed frame.

In order to have a significant value of the intrinsic Schiff moment,
it is not sufficient to have a standard quadrupole deformation: we
need a type of deformation that distinguishes two directions of the
axis violating the symmetry with respect to the reflection in the
equatorial plane perpendicular to the symmetry axis. The collective
effect sought for may be related to the {\sl simultaneous presence
of quadrupole and octupole deformation}, the latter creating a
pear-shaped \cite{auerbach94,spevak95} (or even a heart-shaped
\cite{frauendorf08}) intrinsic mean field. The importance of
octupole deformation for the transmission of statistical parity
violation through intermediate stages of the fission process was
understood long ago \cite{SF82}; this possibility was also
considered in Ref. \cite{FZ95} in relation to the ``sign problem"
(predominance of neutron resonances with the same sign of ${\cal
P}$-violating asymmetry in $^{232}$Th that seemingly contradicts to
the statistical mechanism of enhancement). Now we need the octupole
deformation in the ground state.

In the phenomenological collective description of nuclear deformation in terms
of the equipotential surfaces,
\begin{equation}
R(\theta)=R\left[1+\sum_{l=1}\beta_{l}Y_{l0}(\theta)\right], \label{34}
\end{equation}
the vector terms, $l=1$, emerge, after excluding the center-of-mass
displacement, through bilinear combinations of even and odd multipoles,
\begin{equation}
\beta_{1}=-\sqrt{\frac{27}{4\pi}}\,\sum_{l=2}\frac{l+1}{\sqrt{(2l+1)(2l+3)}}\,
\beta_{l}\beta_{l+1}.                                     \label{35}
\end{equation}
The main contribution that comes from the product of the lowest static
multipoles, quadrupole and octupole, determines the collective intrinsic Schiff
moment \cite{AFS96,SAF97},
\begin{equation}
S_{{\rm intr}}\approx\frac{9}{20\sqrt{35}\pi}\,eZR^{3}\beta_{2}\beta_{3}.
                                                           \label{36}
\end{equation}
The collective character of the octupole moment leads to the strong enhancement
of the intrinsic Schiff moment compared to the single-particle estimates. Of
course, the results are sensitive to the details of the nuclear models, mean
field and effective interactions, but, within a factor of about 2, the Schiff
moment may be enhanced up to two to three orders of magnitude
\cite{AFS96,SAF97,engel03}.

Such results were obtained under an assumption of close levels of opposite
parity mixed by the interaction $W$, with the splitting
$\Delta=|E_{+}-E_{-}|\approx 50$ keV. This is a real situation in $^{225}$Ra
($\Delta=55$ keV, $I=1/2$) and in $^{223}$Ra ($\Delta=50$ keV, $I=3/2$). The
radium and radon isotopes seem to be promising because of clear manifestations
of octupole collectivity. In addition, the large nuclear charge is favorable
for the enhancement of the atomic EDM \cite{flam99}. We need to note that the
resulting space-fixed expectation value of the Schiff moment, according to eqs.
(\ref{33}) and (\ref{36}), is proportional to the product $\alpha S_{{\rm
intr}}$ and therefore to $\beta_{3}^{2}$.

\subsection{Parity doublets}

The mixing can be particularly enhanced if the admixed states are
{\sl parity doublets} \cite{AFS96,SAF97,auerbach94,spevak95,FZ95}.
In the presence of the octupole deformation (or for any axially
symmetric shape with no reflection symmetry in the equatorial
plane), the states of certain parity $\Pi$ are even and odd
combinations of intrinsic states $\chi_{\pm K}$ with the quantum
numbers $\pm K\neq 0$. The intrinsic wave functions which differ
just by the ``right" or ``left" orientation of the pear-shape
configuration should be combined in the states with definite parity
$\Pi$,
\begin{equation}
|IMK;\Pi\rangle=\sqrt{\frac{2I+1}{8\pi}}\left[D^{I}_{MK}\chi_{K}
+\Pi(-)^{I+K}D^{I}_{M-K}\chi_{-K}\right].             \label{37}
\end{equation}
Such doublets in fact do not even require axial symmetry; the label
$\pm K$ may have a more general meaning. The intrinsic partners are
time-conjugate and, according to the Kramers theorem, they are
degenerate in the adiabatic approximation. In the non-axial case,
one can write the wave function as a sum over $K$ of items similar
to those in eq. (\ref{37}).

In reality the doublets (\ref{13}) are split by additional
interactions. This can be accomplished by Coriolis forces (the
body-fixed frame of the rotating nucleus is non-inertial) or by the
tunneling between the two orientations. However such a splitting is
not large and the similarity of intrinsic structure should help in
increasing the mixing by the weak interactions. As explained in
Refs. \cite{AFS96,SAF97,auerbach94,FZ95,SF80}, only the interaction
violating both ${\cal P}$- and ${\cal T}$-invariance can mix the
doublet partners because
\begin{equation}
\langle IMK;-\Pi|W|IMK;\Pi\rangle=
\frac{1}{2}\Bigl[\langle\chi_{K}|W|\chi_{K}\rangle
-\langle\chi_{-K}|W|\chi_{-K}\rangle\Bigr].              \label{38}
\end{equation}
The matrix elements of the pseudoscalar $W$ change sign together
with $K$ which is possible only if the time-reversal invariance is
violated, along with parity. The ``normal" weak interaction is
${\cal T}$-invariant. Therefore it is capable of mixing the parity
doublets only with the help of a mediator, a regular ${\cal
P,T}$-conserving interaction, including that one responsible for the
doublet splitting. This indirect mixing of parity doublets was
suggested in Ref. \cite{FZ95} for explaining the ``sign problem" in
$^{232}$Th. In contrast to this, the ${\cal P,T}$-violating
interaction can mix the parity doublets directly, which is important
for the enhancement of the Schiff moment.

The big enhancement predicted in this situation is illustrated by
the results of calculation given in Ref. \cite{SAF97}, see table
(\ref{39}). We select here the cases where the energy splitting
$\Delta E$ between the partners of the parity doublet is
experimentally known. The parameter $\eta$ is defined in eq.
(\ref{30}).

\begin{equation}
\begin{array}{c|cccccc}
 &^{223}{\rm Ra}&^{225}{\rm Ra}&^{221}{\rm Fr}&^{223}{\rm Fr}&^{225}{\rm
 Ac}&^{229}{\rm Pa}\\
 \hline
 \Delta E_{{\rm exp}}\,({\rm keV})& 50 & 55 & 234 & 161 & 40 & 0.2\\
 S_{{\rm intr}}(e\,{\rm fm}^{3})& 24 & 24 & 21 & 20 & 28 & 25 \\
 S(10^{8}(\eta e\,{\rm fm}^{3})& 400 & 300 & 43 & 500 & 900 & 12 000 \\
 D(10^{25}(\eta e\,{\rm cm}) & 2700 & 2100 & 240 & 2800 & &
                                   \end{array}
                                                       \label{39}
 \end{equation}
The last line contains the results of atomic calculations which can
be in fact extrapolated \cite{SAF97} from the work for lighter
atomic analogs \cite{dzuba85}.

\section{Soft octupole mode}

As was mentioned earlier, in a nucleus with the combination of
developed quadrupole and octupole deformations, the intrinsic Schiff
moment is determined by the collective octupole moment $\beta_{3}$,
whereas the Schiff moment in the space-fixed frame is proportional
to its square. Obviously, the sign of the octupole moment is not
important. This gives rise to the idea \cite{engel00,FZ03} that,
instead of static octupole deformation, the same role of the
enhancing agent can be played by the {\sl dynamic octupole
deformation}.

The soft octupole mode (low-lying collective $3^{-}$ ``one-phonon"
state) is observed in many nuclei and, for a small frequency
$\omega_{3}$ of this mode, the vibrational amplitude increases,
$\langle \beta_{3}^{2}\rangle \propto 1/\omega_{3}$. This relation
would be precise for the harmonic vibrations; its quadrupole analog
sometimes is called the {\sl Grodzins relation} \cite{grodzins68}.
In practice it approximately works although octupole vibrations are
in many cases noticeably fragmented and reveal anharmonicity
\cite{raman91,metlay95}. If the Schiff moment is indeed enhanced
under such conditions without static octupole deformation, this can
provide a more broad choice for the experimental search.
Numerically, the mean square amplitude $\langle
\beta_{3}^{2}\rangle$ is close to the squared value
$\langle\beta_{3}\rangle^{2}$ of static octupole deformation in
pear-shaped nuclei. This value can be extracted from the reduced
transition probability $B({\rm E}3; 0\rightarrow 3^{-})$, see the
compilation in \cite{kibedi02}.

In the presence of the soft octupole mode, the octupole moment
$Q_{3\mu}$ oscillates with the low frequency, and its intrinsic
component along the axis defined by the static quadrupole
deformation $\beta_{2}$ is phenomenologically given by
\begin{equation}
Q_{3}=\frac{3}{4\pi}\,eZR^{3}\beta_{3}.           \label{40}
\end{equation}
This implies, eq. (\ref{36}), the slowly oscillating intrinsic
Schiff moment,
\begin{equation}
S_{{\rm intr}}=\frac{3}{5\sqrt{35}}\,Q_{3}\beta_{2} \label{41}
\end{equation}
(as we have already stressed, the intrinsic Schiff moment does not
depend on violation of fundamental symmetries).

Now we need to introduce the mechanisms converting the intrinsic
Schiff moment into observable ${\cal P,T}$-violating effects. The
description of the previous paragraph referred to the deformed
even-even core. The space-fixed Schiff moment needs the non-zero
nuclear spin so we proceed to the neighboring odd-$A$ nucleus. The
unpaired nucleon interacts with the octupole mode. This dynamic
octupole deformation of the mean field can mix, still in the
body-fixed frame, the single-particle orbitals of opposite parity.
As suggested in Ref. \cite{engel00}, the mixing leads to the
non-vanishing expectation value of the weak interaction $\langle
W\rangle$ in the body-fixed frame. This process can be called {\sl
``particle excitation"}. In a parallel process of {\sl ``core
excitation"} \cite{FZ03}, the octupole component of the weak ${\cal
P,T}$-violating field of the odd particle can excite the soft
octupole mode in the core.

The estimate of the first mechanism can be based on the
octupole-octupole part of the residual nucleon interaction. The
original orbital $|\nu)$ acquires the octupole phonon admixture
while the particle is scattered to some orbitals $|\nu'\rangle$ of
opposite parity,
\begin{equation}
|\nu)\Rightarrow |\tilde{0}\rangle=|\nu;0\rangle+\sum_{\nu'}
a_{\nu'}|\nu';1\rangle,                        \label{42}
\end{equation}
where the number after the semicolon in the state vector indicates
the number of octupole phonons. The orthogonal one-phonon state is,
in the same approximation,
\begin{equation}
|\nu;1\rangle\Rightarrow|\tilde{1}\rangle=|\nu;1\rangle+
\sum_{\nu'}b_{\nu'}|\nu';0\rangle.        \label{43}
\end{equation}
The mixing amplitudes between the orbitals with energies
$\epsilon_{\nu}$ are
\begin{equation}
a_{\nu'}=\frac{\beta_{3}(F_{3})_{\nu'\nu}}{\epsilon_{\nu}
-\epsilon_{\nu'}-\omega_{3}}, \quad
b_{\nu'}=\frac{\beta_{3}(F_{3})_{\nu'\nu}}{\epsilon_{\nu}
-\epsilon_{\nu'}+\omega_{3}},                       \label{44}
\end{equation}
where we assume the octupole forces in the form $\beta_{3}F_{3}$,
the octupole collective coordinate $\beta_{3}$ being defined by eq.
(\ref{40}), while $F_{3}$ is an operator acting on the particle and
having the form close to $-(dU/dr)Y_{30}$ with the radial factor
usually taken as a derivative of the spherical mean field potential,
a reasonable approximation for realistic deformations of low
multipolarities. The quantity $\beta_{3}$ in eq. (\ref{44}) is the
transition matrix element of this collective octupole coordinate
between the ground and one-phonon states in the even-even core.

Now the states $|\tilde{0}\rangle$ and $|\tilde{1}\rangle$ are mixed
by the ${\cal P,T}$-violating potential. This mechanism involves the
coherent part of the weak interaction $\overline{W}$ averaged over
the core nucleons. The mixing matrix element is found as
\begin{equation}
\langle\tilde{0}|\overline{W}|\tilde{1}\rangle=\beta_{3}\sum_{\nu'}\,
\frac{2(\epsilon_{\nu}-\epsilon_{\nu'})}{(\epsilon_{\nu}-
\epsilon_{\nu'})^{2}+\omega_{3}^{2}}\,\overline{W}_{\nu\nu'}
(F_{3})_{\nu'\nu},                                 \label{45}
\end{equation}
which is still an operator linear in the collective coordinate
$\beta_{3}$. In the adiabatic limit, when the octupole mode
frequency $\omega_{3}$ is small compared to the single-particle
spacing between the orbitals of opposite parity, the weak
interaction is essentially acting at a fixed octupole deformation
and then it is averaged over the slowly evolving phonon wave
function. The result practically coincides with that for the static
octupole deformation discussed earlier. The only difference is the
substitution of the static $\beta_{3}^{2}$ by the dynamic mean
square average $\langle\beta_{3}^{2}\rangle$.

In the core excitation mechanism \cite{FZ03}, the effective part of
the weak interaction $W_{ab}$ acts between the valence nucleon $b$
and the paired nucleons $a$ in the core. Because of pairing in the
core, only the contribution proportional to the spin of the valence
nucleon survives,
\begin{equation}
W_{ab}=-\frac{G}{\sqrt{2}}\,\frac{1}{2m}\,\eta_{ba}\Bigl(\nabla_{a}\cdot
\psi^{\dagger}_{b}({\bf r}_{a})\vec{\sigma}_{b}\psi_{b} ({\bf
r}_{a})\Bigr).                               \label{46}
\end{equation}
We need to extract from this interaction the octupole component
$W_{3}$ proportional to the operator $Q_{3}=r^{3}Y_{30}$. The result
\cite{FZ03} depends on the specific orbital of the external nucleon
and can be presented in the form
\begin{equation}
(W_{3})_{a}\approx k\frac{G}{mR^{7}}\,Q_{3}\eta_{ba}   \label{47}
\end{equation}
(this operator has to be multiplied by the creation or annihilation
operator of the $3^{-}$ phonon). Here $k$ is the numerical factor
determined by the spin-orbit structure of the valence orbital; in
typical cases $|k|\approx 0.6$. The matrix element of this
interaction exciting an octupole phonon (that contains both proton
and neutron coherent components) is given by
\begin{equation}
\langle 1|W_{3}|0\rangle=k\frac{G}{mR^{7}}\,AR^{3}\,\frac{3}{4\pi}\,
\langle \beta_{3}\rangle^{2}\eta_{b},                 \label{48}
\end{equation}
where the coupling constant is
$\eta_{b}=(Z/A)\eta_{bp}+(N/A)\eta_{bn}$, and the subscript $b$ is
$n(p)$ for the odd neutron (proton).

Using the mixing produced by the operator $W_{3}$ for calculating
the effective Schiff moment operator and projecting to the
space-fixed frame we come to the result \cite{FZ03} of the same
order of magnitude as in the case of the particle excitation.
Compared to the static octupole deformation, the difference is,
apart from numerical factors of order one, just in the substitution
of static $\beta_{3}^{2}$ by the effective dynamic mean square
value. Taking the limiting value in $^{199}$Hg as a current
standard, we can expect the enhancement in the interval of 100 -
1000 if the energy spacing $\Delta$ is of the order or less than 100
keV. The appropriate candidates are $^{223,225}$Ra, $^{223}$Rn,
$^{223}$Fr, $^{225}$Ac, and maybe $^{239}$Pu, where the estimates of
Ref. \cite{FZ03} are lower than in Ref. \cite{engel00}.

\section{Soft quadrupole and octupole modes}

\subsection{RPA approach}

The results of the previous consideration point out a tempting
possibility of searching for the significant enhancement of the
Schiff moment in a broader class of {\sl spherical} nuclei where
both collective modes, quadrupole and octupole, are clearly
pronounced and have low frequencies. As an example, one can mention
light spherical isotopes of radium and radon. The experimental data
\cite{cocks97} for $^{218,220,222}$Rn and for other even-even nuclei
in this region show long quasivibrational bands of positive and
negative parity, where the energy intervals do not obey the
rotational rules. The phonon frequencies are quite low, and the
strong E1 transitions are observed between the appropriate members
of the quadrupole and octupole bands. The softness of the modes and
large phonon transition probabilities $B({\rm E}2;0\rightarrow
2^{+})$ and $B({\rm E}3;0\rightarrow 3^{-})$, along with strong
dipole interband coupling, indicate that the situation might be
favorable for the enhancement of the Schiff moment.

The mixing of the $2^{+}$ and $3^{-}$ phonons with the valence
particle in a neighboring odd-$A$ nucleus can be considered as a
slow (adiabatic) process of adjustment of the valence orbitals to
the oscillating mean field, as we argued in the previous section. If
the particle can form states with the same spin in both types of
mixing, these states should be rather close in energy and can be
mixed among themselves by the weak interaction. Here we do not
introduce any body-fixed frame so the angular momentum must be
strictly conserved in those mixing processes. Thus, in our main eq.
(\ref{27}), we can have in the odd nucleus states of both parities
with the same $I,M$ quantum numbers like
\begin{equation}
|IM\rangle=\left[C_{0}\alpha^{\dagger}_{jM}\delta_{jI}+\sum_{\lambda
j'}C_{2}(j'\lambda;I)
(\alpha^{\dagger}_{j'}Q^{\dagger}_{\lambda})_{IM} \right]|0\rangle.
                                              \label{49}
\end{equation}
Here $\alpha_{jm}$ and $Q_{\lambda\mu}$ are quasiparticle and phonon
operators, respectively, coupled in the second term of eq.
(\ref{49}) into correct total angular momentum $I$, whereas
$|0\rangle$ represents the ground state of the even nucleus.

The detailed microscopic calculations along these lines were
performed in Ref. \cite{ADFLSZ}. In the neutron-odd nucleus, the
proton contribution needed for the Schiff moment comes from the
transition matrix element of the Schiff operator between the
appropriate states (49) of the same spin $I$ and opposite parity,
\begin{equation}
\langle I^{\pm},M=I|S_{z}|I^{\mp},M=I\rangle=\sum_{\lambda\lambda'j}
X(jI;\lambda\lambda')C_{2}(j\lambda;I^{\pm})C_{2}(j\lambda';I^{\mp})
(\lambda||S||\lambda').                          \label{50}
\end{equation}
where $X(jI;\lambda\lambda')$ are geometric coefficients resulting
from vector coupling of angular momenta. The reduced matrix element
of the Schiff momentum, $(\lambda||S||\lambda')$, is taken between
the phonon states of opposite parity in the even-even core. Because
of the strong dipole coupling between the corresponding bands found
in the candidate nuclei, we expect that this matrix element should
enhance the Schiff moment.

Concrete calculations \cite{ADFLSZ} used the random phase
approximation (RPA) in the form of the quasiparticle-phonon model
\cite{solov}. The multipole-multipole forces are fixed in even
nuclei by the phonon parameters. The result for the Schiff moment
can be expressed in terms of the single-particle Schiff matrix
elements, $(j_{1}||S||j_{2})$, standard pairing amplitudes, $(u,v)$,
and the RPA phonon amplitudes of two-quasiparticle and two-quasihole
components, $(A,B)$,
\[(\lambda||S||\lambda')=\sqrt{35}\sum_{123}(u_{1}u_{2}-v_{1}v_{2})
\left\{\begin{array}{ccc}
\lambda & \lambda' & 1 \\
j_{1} & j_{2} & j_{3} \end{array}\right\}\]
\begin{equation}
\times (j_{1}||S||j_{2})\Bigl[A_{\lambda}(23)A_{\lambda'}(31)+
B_{\lambda}(23)B_{\lambda'}(31)\Bigr].            \label{51}
\end{equation}

The weak interaction was taken in the mean-field form, eq.
(\ref{30}),
\begin{equation}
\overline{W}_{b}({\bf r})=\frac{G}{\sqrt{2}}\,\frac{1}{2m}\, \eta(\vec{\sigma}
\cdot{\bf r})\,\frac{1}{4\pi r}\,\frac{d\rho(r)}{dr},   \label{52}
\end{equation}
where $\rho(r)$ is determined by the pairing occupancy factors in
the core. There are several contributions of the interaction
(\ref{52}) to various parts of the complicated calculation: wave
functions of the unpaired quasiparticle, matrix elements of
quasiparticle-phonon coupling, intermediate particle and phonon
propagators, and phonon loops. Combining these calculations with the
energy denominators we come to the final results.

At this stage we could not find an enhancement of the nuclear Schiff
moment. For example, for the $^{219}$Rn isotope the matrix element
of the weak interaction equals -1.3 $\eta\cdot 10^{-2}$ eV, and the
final value of the ground state Schiff moment was 0.30 $\eta\cdot
10^{-8}e\cdot$fm$^{3}$. Typically, the reduced matrix elements
$(2^{+}|S|3^{-})$ in the even nucleus are of the order (1-2)
$e\cdot$fm$^{3}$, and the matrix elements of the Schiff operator
between the ground state in the odd nucleus and its parity partner
are around 0.1-0.2 $e\cdot$fm$^{3}$. Final results for the Schiff
moment are of the same order as in pure single-particle models (the
single-particle contribution unrelated to the soft modes
\cite{FKS86,DS03,DSA05} has to be added).

These calculations seemingly contradict the idea of a possible enhancement by
soft collective modes. Nevertheless, a useful exercise \cite{ADFLSZ} confirms
that the effect indeed exists but, in the RPA framework, requires artificially
low collective frequencies when the dynamic deformation amplitudes increase as
$\beta\propto 1/\omega$. One can consider the theoretical RPA limit of
collapsing frequencies,
\begin{equation}
\omega_{2,3}\Rightarrow y\omega_{2,3}, \quad y\ll 1,      \label{53}
\end{equation}
and accurately separate the singular part of the RPA solutions. As the
collective frequencies go down, the reduced matrix element $(2^{+}|S|3^{-})$ in
the even nucleus, the mixing matrix element of the weak interaction in the odd
nucleus and the final Schiff moment grow large. These trends are seen in the
following table (\ref{54}),
\begin{equation}
\begin{array}{c|c|c|c|c|c}
{\rm Nucleus} & y & (2^{+}|S|3^{-}) & {\rm m.e.}\;W & {\rm m.e.}\;S & S\\
\hline
^{219}{\rm Ra} &1& 1.7 & -1.3 & -0.1 & 0.3 \\
&              0.1& 20 & 1.1 & -0.2 & -0.2 \\
&            0.01 & 195 & 53 & -0.2 & 6.2 \\
\hline
^{221}{\rm Ra} & 1 & 2.2 & 0.2 & -0.2 & -0.1\\
 &              0.1 & 23 & -19 & -0.5 & 6 \\
 &            0.01 & 235 & -253 & -2.7 & 560 \end{array}
                                                      \label{54}
\end{equation}
It is clear from the table that the matrix element $(2^{+}|S|3^{-})$ indeed
increases $\propto 1/\omega$. Other matrix elements are also sensitive to the
level spacing in the odd nucleus. Here we need to mention that the RPA results
with the parameters fitted to the phonon frequencies do not produce a
satisfactory description of entire spectra in odd nuclei.

\subsection{Going beyond RPA}

To summarize the findings of the previous subsection, in the situation when the
phonon-quasiparticle coupling becomes strong, the standard RPA approach that
accounts for a single-phonon admixture to quasiparticle wave functions, is not
adequate. The effect of enhancement appears either with static deformation or
in the strong coupling limit when effectively the condensate of phonons emerges
that mimics the deformed field. In the exactly solvable particle-core model
\cite{BZ65} with the soft {\sl monopole} mode, $\lambda=0$, the ground state of
the odd-$A$ nucleus contains a coherent phonon state with the average number of
phonons defined by the coupling constant. The quasiparticle strength in this
regime is strongly fragmented over many excited states. Similar effects should
take place for quadrupole and octupole modes \cite{BM52,abbas81,stoyanov04}
when the coherence finally leads to the phase transition to static deformation.

In agreement with above arguments, the calculations \cite{ADFLSZ}
with artificially quenched frequencies show that the wave function
of the odd nucleus becomes exceedingly fragmented. For example, in
the realistic case, $y=1$, for the ground state $I=7/2$ in
$^{219}$Ra, there exists only one large combination of amplitudes
required for the mixing, namely there are particle-phonon states
$7/2^{+}$ with the wave function $(2g_{9/2},2^{+})_{7/2}$ and
$7/2^{-}$ with the wave function $(2g_{9/2},3^{-})_{7/2}$; their
weights in the full RPA wave functions are 98\% for negative parity
but only 8\% for positive parity. With quenching of frequencies,
these amplitudes are getting drastically reduced, up to 2\% for
negative parity and 1\% for positive parity. Only after the
spreading of the single-particle strength reached saturation in the
orbital space under consideration, one can indeed see the
enhancement of the Schiff moment.

Thus, the conventional RPA ansatz for the wave function of the odd nucleus as a
superposition of particle+phonon components is invalid under conditions of soft
collective modes. Many-phonon components take over a large fraction of the
total wave function. Moreover, soft modes become mutually correlated. The
correlation between soft quadrupole and octupole excitations was suggested in
the global review of octupole vibrations \cite{metlay95}. The presence of the
octupole phonon singles out the intrinsic axis (similar to the static
deformation illustrated by Fig. 1) and triggers the {\sl spontaneous breaking
of rotational symmetry} with an effective quadrupole condensate emerging. This
mechanism follows from the simplest construction of the phenomenological
coupling between the octupole and quadrupole modes that accounts for parity and
angular momentum conservation, The effective Hamiltonian of this type is given
by
\begin{equation}
H=H_{2}+H_{3}+H_{23},                             \label{55}
\end{equation}
where $H_{2}$ and $H_{3}$ describe quadrupole and octupole
collective modes (in principle including their anharmonicity). The
interaction described by the destruction and creation operators
$d_{\mu},d^{\dagger}_{\mu}$ and $f_{\mu},f^{\dagger}_{\mu}$, for the
quadrupole and octupole phonons, respectively,
\begin{equation}
H_{23}=x\sum_{\mu}\Bigl[(f^{\dagger}f)_{2\mu}d_{\mu}^{\dagger}+ {\rm
h.c.}\Bigr],                                   \label{56}
\end{equation}
generates the equation of motion
\begin{equation}
[d_{\mu},H]=\omega_{2}d_{\mu}+x(f^{\dagger}f)_{2\mu},   \label{57}
\end{equation}
where $\omega_{2}$ is the quadrupole frequency and $x$ the coupling
constant. In the presence of the octupole phonon, we obtain the
quadrupole condensate,
\begin{equation}
\langle d_{\mu}\rangle=-\frac{x}{\omega_{2}}\,
\langle(f^{\dagger}f)_{2\mu}\rangle,              \label{58}
\end{equation}
with the coherent intensity inversely proportional to the frequency
as a characteristic feature of the situation associated with soft
modes.

The direction of the effective self-consistent deformation is
arbitrary, and, to restore the symmetry and appropriate quantum
numbers of total nuclear spin in the space-fixed coordinate frame,
we accept that the orientation of the deformation is given by a
spherical function of corresponding rank,
\begin{equation}
\langle d_{\mu}\rangle=\delta_{2}\,\sqrt{\frac{4\pi}{5}}\,
Y_{2\mu}^{\ast}({\bf n}),\quad\langle f_{\mu}\rangle=\hat{f}\,
\sqrt{\frac{4\pi}{7}}\, Y_{3\mu}^{\ast}({\bf n}).   \label{59}
\end{equation}
Here ${\bf n}$ is the unit vector of the symmetry axis considered as
a variable in the collective space. A similar operator approach was
used long ago in the derivation of the nuclear moment of inertia
without applying a cranking model \citep{BZ70}. The number
$N_{3}=\sum_{\mu}f_{\mu}^{\dagger}f_{\mu}= \hat{f}^{\dagger}\hat{f}$
of octupole phonons is conserved by the Hamiltonian (\ref{56}),
$N_{3}=1$ in the lowest $3^{-}$ state, and $\hat{f}^{\dagger}$ is
the operator generating the octupole vibrational mode in the
body-fixed frame defined by the orientation ${\bf n}$. Then eq.
(\ref{58}) equates the ${\bf n}$-dependence and, with the ansatz
(\ref{59}), provides the effective quadrupole deformation parameter
$\delta_{2}$,
\begin{equation}
\delta_{2}=-\frac{x}{\omega_{2}}\,\sqrt{5}\left(\begin{array}{ccc}
3 & 3 & 2\\
0 & 0 &
0\end{array}\right)=-\sqrt{\frac{4}{21}}\,\frac{x}{\omega_{2}}.
                                                     \label{60}
\end{equation}

The equation of motion for the octupole mode, given by the
commutator $[f_{\mu},H]$ and the effective quadrupole parameter
(\ref{60}), is linear. It relates the excitation energy $E_{3}$ of
the octupole phonon with the corresponding unperturbed energy
$\omega_{3}$ and the quadrupole condensate (\ref{60}). Collecting
again the terms expressing the angular dependence, we obtain
\begin{equation}
E_{3}=\omega_{3}-\frac{8}{21}\,\frac{x^{2}}{\omega_{2}}.\label{61}
\end{equation}
This simple regularity \cite{metlay95} provides a clear correlation
between the two modes. Recent measurements for the long chain of
even-even xenon isotopes \citep{mueller06} show precisely such a
correlation, with a rather large magnitude for the parameter $x$
that exceeds the expectations for the anharmonic mode-mode coupling
based on the standard RPA estimates.

A similar effect of condensation is brought in by the odd particle.
The effective particle-phonon interaction is linear in phonon
operators and proportional to different components of the particle
density matrix
\begin{equation}
\rho_{j_{1}m_{1};j_{2}m_{2}}=\langle
a_{j_{2}m_{2}}^{\dagger}a_{j_{1}m_{1}}\rangle=
\sum_{L\Lambda}(-)^{L-\Lambda+j_{2}-m_{2}}\left(\begin{array}{ccc}
j_{1} & L & j_{2}\\
m_{1} & -\Lambda & -m_{2}\end{array}\right)\rho_{L}(j_{1}j_{2})
Y_{L\Lambda}^{\ast}({\bf n}).                  \label{62}
\end{equation}
in terms of the operators of creation, $a_{jm}^{\dagger}$, and annihilation,
$a_{jm}$, of the particle in the spherical basis. Here the even-$L$ parts come
from the pairs of levels $(j_{1},j_{2})$ or $(j_{1}',j_{2}')$ of the same
parity, whereas the odd-$L$ ones correspond to the combinations
$(j_{1},j_{2}')$ and $(j_{1}',j_{2})$ of single-particle levels of opposite
parity. The equations of motion for phonons in the odd nucleus bring in their
coherent states signaling the onset of an effective deformation, now for both,
quadrupole and octupole modes. Self-consistently, the unpaired particle
occupies the Nillson-type orbitals in the deformed field characterized by the
parameters $\beta_{2}$ and $\beta_{3}$ determined by the coupling constants
with the particle and proportional to $1/\omega_{2}$ and $1/\omega_{3}$,
respectively.

The operator $S_{1\nu}$ of the Schiff moment in the even nucleus has
a reduced matrix element $S^{\circ}\equiv(2||S_{1}||3)$ between the
low-lying $2^{+}$ and $3^{-}$ states. As mentioned earlier, the
dipole transitions between the states of the quadrupole and octupole
bands are empirically known to be enhanced in nuclei of our
interest, such as light radium and radon isotopes \citep{cocks97}.
The collective contribution to this operator can be written in terms
of our phonon variables as
\begin{equation}
S_{1\nu}=S^{\circ}\sum_{\mu\mu'}(-)^{\nu+\mu}
\left(\begin{array}{ccc}
1 & 2 & 3\\
-\nu & -\mu & \mu'\end{array}\right)
\Bigl(d_{\mu}^{\dagger}f_{\mu'}+
(-)^{\mu+\mu'}f_{-\mu'}^{\dagger}d_{-\mu}\Bigr).       \label{63}
\end{equation}
With the ground state expectation values of the effective
deformation parameters in the odd nucleus, this gives a rotational
operator
\begin{equation}
\frac{S_{1\nu}}{S^{\circ}}=-\frac{1}{\sqrt{\pi}}\,\beta_{2}\beta_{3}
Y_{1\nu}^{\ast}                               \label{64}
\end{equation}
enhanced by small collective frequencies. As a result, we reduce the whole
problem to that of the {\sl particle + rotor} type \cite{leander84,SAF97},
where the static deformation is substituted by the effective deformation coming
from the soft quadrupole and octupole modes of the spherical even core. The
observable Schiff moment in the laboratory frame can come only from explicitly
acting with the ${\cal P}$- and ${\cal T}$-violating weak interaction $W$ that
creates an admixture of the states $|n\rangle$ having opposite parity to the
ground state $|0\rangle$ but the same spin $I$.

In this spirit, a model of two single-particle levels of the same large $j$ and
opposite parity with $n$ particles interacting through pairing and
multipole-multipole (quadrupole and octupole) forces in the presence of the
${\cal P,T}$-violating weak interaction was considered using the exact
diagonalization instead of the RPA \cite{ZVA}. The model does not introduce the
intrinsic frame and preserves exact quantum constants of motion at all stages
of calculation. The orbital space of the model is not large enough to
demonstrate the constructive mutual support of the two soft modes; instead they
compete for available quasiparticle excitations. Nevertheless, the model
reveals the existence of a parameter region, where both frequencies in the
even-even system are sufficiently low, while, in the neighboring odd system,
matrix elements of the weak interaction and of the Schiff moment are
significantly enhanced. One also clearly sees the appearance of parity doublets
at a small spacing in the same parameter region. This set of conditions is
favorable for a strong enhancement of the expectation value of the Schiff
moment. The realistic calculations in the future will show how reliable is this
evaluation.

\subsection{Nuclear structure aspects}

The consideration of mechanisms responsible for possible enhancement of
violation of fundamental symmetries in many-body systems, such as atomic
nuclei, elucidates also the problems of our understanding of nuclear structure
and our ability to develop corresponding realistic theories. The statistical
mechanism related to the uniform structure of compound states at high level
density is in general understood. Even if it is impossible to calculate the
properties of each individual state, we have sufficient knowledge of typical
regularities, and in the region of quantum many-body chaos all states within a
certain energy window ``look the same" \cite{percival73,big}. The situation is
different near the ground state, in the region important for the search of the
atomic EDM.

As we have tried to argue, the low-lying collective modes and strong
interaction between them and with quasiparticle excitations may lead to the
strong enhancement of the effects we are interested in. Unfortunately, current
microscopic nuclear structure theory does not give a clear answer to the
question of the realistic strength of required interactions and their
compatibility with the standard picture of nuclear shells. The two main
obstacles in this direction are the necessity of large orbital space (we are
looking at heavy nuclei) and the absence of reliable effective interactions,
although the first feature supposedly can be treated with the aid of the
exponential convergence method \cite{HVZ99,HBZ02} or similar approaches. The
lack of knowledge of interactions requires more deep studies.

We have seen the possible vital role of the three-phonon couplings, as in eq.
(\ref{51}), in the development of enhancement. In the conventional RPA
framework, such couplings are expressed by triangular diagrams, which come with
a considerable reduction due to the combinations $u_{1}u_{2}-v_{1}v_{2}$ of the
pairing coherence factors. This combination is antisymmetric with respect to
the single-particle occupancies and would vanish in the case of full symmetry
around the Fermi surface. This can be understood in analogy with the well known
{\sl Furry theorem} of quantum electrodynamics. In QED, such three-photon
diagrams vanish exactly because of the precise cancellation of electron and
positron contributions to the loop with three photon tails. In the discrete
nuclear spectrum, there is no full symmetry and the result does not vanish but
still it is considerably suppressed. Some systematic features of interplay
between quadrupole and octupole degrees of freedom, as for example found in
Refs. \cite{metlay95,mueller06}, indicate that the three-phonon vertices should
be stronger that it comes from the RPA estimates.

This fact may have something to do with the effects of three-body
forces whose role in many-body dynamics essentially is unknown. At
this point it makes no difference what is the source of these
forces; they can come from bare nucleon interactions
\cite{schwenk08} or effectively result from the medium
modifications. In the search for three-body interactions in heavy
nuclei it would be natural to start looking for their {\sl
collective} effects \cite{dallas06}. Such effects of cubic
anharmonicity should be visible also in shape phase transitions in
heavy nuclei. Indeed, the cubic quadrupole term in the collective
potential energy, $\sim\beta^{3}\cos(3\gamma)$ is responsible for
sharp restructuring of single-particle orbitals and transition from
gamma-unstable configurations typical for soft vibrators to the well
deformed rotors. This topic definitely deserves further development.

\section{Conclusion}

In this short review we tried to demonstrate the abundance of ideas
and physical images related to the search of the effects of ${\cal
P,T}$- violating forces in atomic nuclei. Of course, there is
immediate interest in measuring such effects which would lead us
beyond the Standard Model, while currently we know only the upper
limits. Because of extreme difficulty of such experiments and their
time-consuming nature, it is important to try to establish the most
promising path and to select nuclei where we can expect the most
pronounced effects.

Along with that, it turns out that the wealth of physics related to
the violation of fundamental symmetries in nuclei elucidates also
many particular problems of nuclear structure which until now did
not have definite answers. These problems are related to various
manifestations of quantum-mechanical symmetries in a strongly
interacting self-bound many-body system, such as the complex
nucleus.  Another open question is that of strong interaction
between various collective and single-particle degrees of freedom.

Parity violation is known to be enhanced by the orders of magnitude by
statistical (chaotic) properties of compound state neutron resonances. In the
search for the ${\cal P,T}$-violation we are looking for coherent effects. The
EDM of the atoms is induced by the nuclear Schiff moment through its ${\cal
P,T}$-violating potential. The best perspectives for a significant enhancement
of the nuclear Schiff moment are currently seen in the nuclei with static
octupole deformation in the ground state. We argued that the soft octupole mode
in a combination with well developed quadrupole deformation is expected to
display enhancement as well. Finally, we came to soft nuclei with slow
quadrupole {\sl and} octupole motion of large amplitude. Although the direct
attempt in this direction did not yet bring desired results, we need to better
understand nuclear physics of such nuclei where the shape is in fact
ill-defined and the routine theoretical methods, such as the RPA, are probably
not sufficient. This leads to new problems of structure of mesoscopic systems
on the verge of shape instability. Another interesting question is that of the
three-body residual forces (coming from bare three-body forces or induced by
the nucleon correlations). Such forces may give stronger mode-mode coupling not
limited by the Furry theorem discussed above. In general, the entire area of
research is very promising for understanding the fundamental symmetries at work
in a many-body environment.

\section{Acknowledgements}

The work reviewed in this article was supported by the NSF grants
PHY-0244453 and PHY-0555366, and by the grant from the Binational
Science Foundation US-Israel. We also acknowledge support from the
National Superconducting Cyclotron Laboratory at the Michigan State
University and from the Institute for Nuclear Theory at the
University of Washington. We thank V.F. Dmitriev, V.V. Flambaum,
R.A. Sen'kov and A. Volya for collaboration and many illuminating
discussions.


\begin{thebibliography}{99}

\bibitem{wu57} C.S. Wu, E. Ambler, R.W. Hayward, D.D. Hoppes, and
R.F. Hudson, Phys. Rev. {\bf 105}, 1413 (1957).

\bibitem{leeyang56} T.D. Lee and C.N. Yang, Phys. Rev. {\bf 104},
254 (1956).

\bibitem{horowitz01}  C.J. Horowitz, S.J. Pollock, P.A. Souder,
and R. Michaels, Phys. Rev. C {\bf 63}, 025501 (2001).

\bibitem{brown00} B.A. Brown, Phys. Rev. Lett. {\bf 85}, 5296 (2000).

\bibitem{piekarewicz08} J. Piekarewicz, nucl-th. arXiv:0802.4029.

\bibitem{erler05} J. Erler and M.J. Ramsey-Musolf, Prog. Nucl. Part. Phys.
{\bf 54}, 351 (2005).

\bibitem{sakharov67} A.D. Sakharov, Pisma Zh. Eksp. Teor. Fiz. {\bf 5}, 32
(1967) [JETP Lett. {\bf 5}, 24 (1967)].

\bibitem{fleischer02} R. Fleischer, Phys. Rep. {\bf 370}, 537
(2002).

\bibitem{ginges04} J.S.M. Ginges and V.V. Flambaum, Phys. Rep.
{\bf 397}, 63 (2004).

\bibitem{alfimenkov83} V.P. Alfimenkov, S.V. Borzakov, V. Van Thuan,
Yu.D. Mareev, L.B. Pikelner, A.Z. Khrykin, and E.I. Sharapov, Nucl.
Phys. {\bf A398}, 93 (1983).

\bibitem{bowman93} J.D. Bowman, G.T. Garvey, M.B. Johnson, and G.E.
Mitchell, Ann. Rev. Nucl. Part. Sci. {\bf 43}, 829 (1993).

\bibitem{frankle93} C.M. Frankle, S.I. Seestrom, N.R. Robertson,
Yu.P. Popov, and E.I. Sharapov, Phys. Part. Nucl. {\bf 24}, 401
(1993).

\bibitem{flamgrib95} V.V. Flambaum and G.F. Gribakin, Progr. Part.
Nucl. Phys. {\bf 35}, 423 (1995).

\bibitem{mitchell99} G.E. Mitchell, J.D. Bowman, and H.A.
Weidenm\"{u}ller, Rev. Mod. Phys. {\bf 71}, 445 (1999).

\bibitem{blin60} R. J. Blin-Stoyle, Phys. Rev. {\bf 120}, 181 (1960).

\bibitem{SF80} O.P. Sushkov and V.V. Flambaum, Pis'ma Zh. Eksp.
Teor. Fiz. {\bf 32}, 377 (1980) [JETP Lett. {\bf 32}, 352 (1980)].

\bibitem{SF82} O.P. Sushkov and V.V. Flambaum, Usp. Fiz. Nauk {\bf
136}, 3 (1982) [Sov. Phys. Usp. {\bf 25}, 1 (1982)].

\bibitem{ABowman92} N. Auerbach and J.D. Bowman, Phys. Rev. C {\bf 46}, 2582
(1992).

\bibitem{ABAB94} N. Auerbach and B.A. Brown, Phys. Lett. B {\bf 340}, 6 (1994).

\bibitem{danilyan80} G.V. Danilyan, Sov. Phys. Usp. {\bf 23} (1980)
323.

\bibitem{big} V. Zelevinsky, B.A. Brown, N. Frazier, and M. Horoi,
Phys. Rep. {\bf 276}, 85 (1996).

\bibitem{ann} V. Zelevinsky, Ann. Rev. Nucl. Part. Sci., {\bf 46},
237 (1996).

\bibitem{LLQM} L.D. landau and E.M. Lifshitz, {\sl Quantum Mechanics
(Non-relativistic Theory)} (Butterworth-Heinemann, Oxford, 1981).

\bibitem{kotzle00} A. K\"{o}tzle, P. Jesinger, F. Gönnenwein, G. A.
Petrov, V. I. Petrova, A. M. Gagarsky, G. Danilyan, O. Zimmer and V.
V. Nesvizhevsky. Nucl. Instr. Meth. A {\bf 440}, 750 (2000).

\bibitem{blinbook} R.J. Blin-Stoyle, {\sl Fundamental Interactions and the
Nucleus} (North-Holland, Amsterdam, 1973).

\bibitem{french88} J.B. French, V.K.B. Kota, A. Pandey, and S.
Tomsovic, Ann. Phys. (N.Y.) {\bf 181}, 198, 235 (1988).

\bibitem{purcell50} E.M. Purcell and N.F. Ramsey, Phys. Rev. {\bf 78},
807 (1950).

\bibitem{landau57} L.D. Landau, Sov. Phys. JETP {\bf 5}, 336 (1957).

\bibitem{jacobs95} J.P. Jacobs, W.M. Klipstein, S.K. Lamoreaux, B.R.
Heckel, and E.N. Fortson, Phys. Rev. A {\bf 52}, 3521 (1995).

\bibitem{romalis01} M.V. Romalis, W.C. Griffith, J.P. Jacobs, and
E.N. Fortson, Phys. Rev. Lett. {\bf 86}, 2505 (2001).

\bibitem{haxton83} W.C. Haxton and E.M. Henley, Phys. Rev. Lett. {\bf 51},
1937 (1983).

\bibitem{SAF97} V. Spevak, N. Auerbach, and V.V. Flambaum, Phys.
Rev. C {\bf 56}, 1357 (1997).

\bibitem{reexamschiff} R.A. Sen'kov, N. Auerbach, V.V. Flambaum, and
V.G. Zelevinsky, Phys. Rev. A {\bf 77}, 014101 (2008).

\bibitem{schiff63} L.I. Schiff, Phys. Rev. {\bf 132}, 2194 (1963).

\bibitem{sandars67} P.G.H. Sandars, Phys. Rev. Lett. {\bf 19}, 1396
(1967).

\bibitem{liu07} C.-P. Liu, M.J. Ramsey-Musolf, W.C. Haxton, R.G.E.
Timmermans, and A.E.L. Dieperink, Phys. Rev. C {\bf 76}, 035503
(2007).

\bibitem{dzuba07} V.A. Dzuba, V.V. Flambaum, and J.S.M. Ginges,
Phys. Rev. A {\bf 76}, 034501 (2007).

\bibitem{FKS84} V.V. Flambaum, I.B. Khriplovich, and O.P. Sushkov,
Zh. Eksp. Teor. Fiz. {\bf 87}, 1521 (1984) [JETP {\bf 60}, 873 (1984)].

\bibitem{flamginges02} V.V. Flambaum and J.S.M. Ginges, Phys. Rev. A
{\bf 65}, 032113 (2002).

\bibitem{dmitriev05} V.F. Dmitrriev and V.V. Flambaum, Phys. Rev. C
{\bf 71}, 068501 (2005).

\bibitem{harakeh81} M.N. Harakeh and A.E.L. Dieperink, Phys. Rev. C
{\bf 23}, 2329 (1981).

\bibitem{khatsymovsky88} V.M. Khatsymovsky, I.B. Khriplovich, and
A.S. Yelkhovsky, Ann. Phys. (N.Y.) {\bf 186}, 1 (1988).

\bibitem{falk99} T. Falk, K.A. Olive, M. Pospelov, and R. Roiban,
Nucl. Phys. {\bf B560}, 3 (1999).

\bibitem{dzuba02} V.A. Dzuba, V.V. Flambaum, J.S.M. Ginges,
and M.G. Kozlov, Phys. Rev. A {\bf 66}, 012111 (2002).

\bibitem{khriplovich98} I.B. Khriplovich, Phys. Lett. B {\bf 444},
98 (1998).

\bibitem{farley04} F.J. Farley, K. Jungmann, J.P. Miller, W.M. Morse,
Y.F. Orlov, B.L. Roberts, Y.K. Semertzidis, A. Silenko, and E.J.
Stephenson, Phys. Rev. Lett. 93, 052001 (2004).

\bibitem{orlov06} Y.F. Orlov, W.M. Morse, and Y.K. Semertzidis,
Phys.Rev.Lett. 96, 214802 (2006).

\bibitem{rosenberry01} M.A. Rosenberry and T.E. Chupp, Phys. Rev.
Lett. {\bf 86}, 22 (2001).

\bibitem{lamoreaux87} S.K. Lamoreaux, J.P. Jacobs, B.R. Heckel, F.J.
Raab, and E. Fortson, Phys. Rev. Lett. {\bf 59}, 2275 (1987).

\bibitem{regan02} B.C. Regan, E.D. Commins, C.J. Schmidt, and D.
DeMille, Phys. Rev. Lett. {\bf 88}, 071805 (2002).

\bibitem{commins94} E.D. Commins, S.B. Ross, D. DeMille, and B.C.
Regan, Phys. Rev. A {\bf 50}, 2960 (1994).

\bibitem{dzuba00} V.A. Dzuba, V.V. Flambaum, and J.S.M. Ginges,
Phys. Rev. A {\bf 61}, 062509 (2000).

\bibitem{guest07} J.R. Guest, N.D. Scielzo, I. Ahmad, K. Bailey,
J.P. Greene, R.J. Holt, Z.-T. Lu, T.P. O'Connor, and D.H.
Potterveld, Phys. Rev. Lett. {\bf 98}, 093001 (2007).

\bibitem{kitano88} M. Kitano {\sl et al.}, Phys. Rev. Lett. {\bf
60},  2133 (1988).

\bibitem{INT07} The 4th ANL/MSU/INT/JINA RIA Theory Workshop, Sept.
19-22, 2007, INT, Seattle (talks online).

\bibitem{hudson02} J.J. Hudson, B.E. Sauer, M.R. Tarbutt, and E.A. Hinds,
Phys. Rev. Lett. {\bf 89}, 023003 (2002).

\bibitem{lamoreaux02} S.K. Lamoreaux, Phys. Rev. A {\bf 66}, 022109
(2002).

\bibitem{mukhamedjanov05} T.N. Mukhamedjanov and O.P. Sushkov,
Phys. Rev. A {\bf 72}, 034501 (2005).

\bibitem{BM} A. Bohr and B. Mottelson, {\sl Nuclear Structure} (Benjamin, New
York, 1975), vol. 2.

\bibitem{FV94} V.V. Flambaum and O.K. Vorov, Phys. Rev. C {\bf 49}, R1827
(1994).

\bibitem{FKS86} V.V. Flambaum, I.B. Khriplovich and O.P. Sushkov,
Nucl. Phys. {\bf A449}, 750 (1986).

\bibitem{DS03} V.F. Dmitriev and R.A. Sen'kov, Yad. Fiz. {\bf 66}, 1988
(2003) [Phys. At. Nucl. {\bf 66}, 1940 (2003)].

\bibitem{DSA05} V.F. Dmitriev, R.A. Sen'kov, and N. Auerbach, Phys. Rev. C
{\bf 71}, 035501 (2005).

\bibitem{jesus05} J.H. de Jesus and J. Engel, Phys. Rev. C {\bf 72},
045503 (2005).

\bibitem{auerbach94} N. Auerbach, Nucl. Phys. {\bf A577}, 443c
(1994).

\bibitem{spevak95} V.Spevak and N.Auerbach, Phys.Lett. B {\bf 359}, 254 (1995).

\bibitem{frauendorf08} S. Frauendorf, Phys. Rev. C {\bf 77},
021304(R) (2008).

\bibitem{FZ95} V.V. Flambaum and V.G. Zelevinsky, Phys. Lett. B {\bf 350},
8 (1995).

\bibitem{AFS96} N. Auerbach, V.V. Flambaum, and V. Spevak,
Phys. Rev. Lett. {\bf 76}, 4316 (1996).

\bibitem{engel03} J. Engel, M. Bender, J. Dobaczewski, J.H. de Jesus, and
P. Olbratowski, Phys. Rev. C {\bf 68}, 025501 (2003).

\bibitem{flam99} V.V. Flambaum, Phys. Rev. A {\bf 60}, R2611 (1999).

\bibitem{dzuba85} V.A. Dzuba, V.V. Flambaum, and P.G. Silvestrov,
Phys. Lett. B {\bf 154}, 93 (1985).

\bibitem{FZ03} V.V. Flambaum and V.G. Zelevinsky, Phys. Rev. C {\bf 68},
035502 (2003).

\bibitem{grodzins68} L. Grodzins, Ann. Rev. Nucl. Sci. {\bf 18},
291 (1968).

\bibitem{raman91} S. Raman, C.W. Nestor, Jr., S. Kahane, and K.H. Bhatt,
Phys. Rev. C {\bf 43}, 556 (1991).

\bibitem{metlay95} M.P. Metlay, J.L. Johnson, J.D. Canterbury,
P.D. Cottle, C.W. Nestor, S. Raman, and V.G. Zelevinsky, Phys. Rev.
C {\bf 52}, 1801 (1995).

\bibitem{kibedi02} T. Kib$\acute{e}$di and R.H. Spear, Atomic Data
and Nuclear Data Tables, {\bf 80}, 35 (2002).

\bibitem{engel00} J. Engel, J.L. Friar, and A.C. Hayes, Phys. Rev. C
{\bf 61}, 035502 (2000).

\bibitem{cocks97} J.F.C. Cocks {\sl et al.}, Phys. Rev. Lett. {\bf 78},
2920 (1997).

\bibitem{ADFLSZ} N. Auerbach, V.F. Dmitriev, V.V. Flambaum, A. Lisetsky,
R.A. Sen'kov, and V.G. Zelevinsky, Phys. Rev. C {\bf 74}, 025502
(2006).

\bibitem{solov} V.G. Soloviev, {\sl Theory of Atomic
Nuclei: Quasiparticles and Phonons} (IOP Publishing, Bristol, 1992).

\bibitem{BZ65} S.T. Belyaev and V.G. Zelevinsky, Yad. Fiz. {\bf 2}, 615
(1965) [Sov. J. Nucl. Phys. {\bf 2}, 442 (1966)].

\bibitem{BM52} A. Bohr, Mat. Fys. Medd. Dan. Vidensk. Selsk. {\bf 26},
\#14 (1952); A. Bohr and B.R. Mottelson, {\sl ibid.} {\bf 27}, \#16
(1953).

\bibitem{abbas81} A. Abbas, N. Auerbach, N. Van Giai, and L. Zamick,
Nucl. Phys. {\bf A367}, 189 (1981).

\bibitem{stoyanov04} Ch. Stoyanov and V. Zelevinsky, Phys. Rev. C
{\bf 70}, 014302 (2004).

\bibitem{mueller06} W.F. Mueller {\sl et al.,} Phys. Rev. C {\bf 73},
014316 (2006).

\bibitem{BZ70} S.T. Belyaev and V.G. Zelevinsky, Yad. Phys. {\bf 11},
741 (1970) {[}Sov. J. Nucl. Phys. {\bf 11}, 416 (1970)].

\bibitem{leander84} G.A. Leander and R.K. Sheline, Nucl. Phys. {\bf A413},
375 (1984).

\bibitem{ZVA} V. Zelevinsky, A. Volya, and N. Auerbach,
nucl-th/0803.2915.

\bibitem{percival73} I.C. Percival, J. Phys. {\bf B6}, L229 (1973).

\bibitem{HVZ99} M. Horoi, A. Volya and V. Zelevinsky, Phys. Rev.
Lett. {\bf 82}, 2064 (1999).

\bibitem{HBZ02} M. Horoi, B.A. Brown and V. Zelevinsky, Phys. Rev.
C {\bf 65}, 027303 (2002).

\bibitem{schwenk08} A. Schwenk and J.D. Holt, arXiv:0802.3741[nucl-th],
2008.

\bibitem{dallas06} V. Zelevinsky, T. Sumaryada, and A. Volya,
http://meeting.aps.org/link/BAPS.2006.APR.C8.4.


\end{thebibliography}
\end{document}